\documentclass[apj]{emulateapj}
\usepackage{apjfonts} 
\usepackage[hang,flushmargin]{footmisc}
\usepackage{amsmath,amstext}
\usepackage[breaklinks,colorlinks,citecolor=blue,linkcolor=magenta, urlcolor=red]{hyperref} 
\usepackage[all]{hypcap} 
\usepackage{afterpage}
\usepackage[export]{adjustbox}
\usepackage{soul}

\newcommand{\JWST}{{\sl JWST }}
\newcommand{\Spitzer}{{\sl Spitzer }}

\usepackage{lipsum} 

\slugcomment{Accepted for Publication in AJ December 17, 2019}

\shorttitle{Evaluating Climate Variability of the Canonical Hot Jupiters
HD 189733\MakeLowercase{b} \& HD 209458\MakeLowercase{b}}
\shortauthors{Kilpatrick et al.}

\begin{document}

   \title{Evaluating Climate Variability of the Canonical Hot Jupiters
HD 189733\MakeLowercase{b} \& HD 209458\MakeLowercase{b}\\
Through Multi-Epoch Eclipse Observations.  }

   \author{Brian M. Kilpatrick\altaffilmark{1, 2,*},
    Tiffany Kataria\altaffilmark{3},
  Nikole K. Lewis\altaffilmark{4},
  Robert T. Zellem\altaffilmark{3},
  Gregory W. Henry\altaffilmark{5},\\
  Nicolas B. Cowan\altaffilmark{6},
  Julien de Wit\altaffilmark{7},
  Jonathan J. Fortney\altaffilmark{8},
  Heather Knutson\altaffilmark{9}, 
  Sara Seager\altaffilmark{7},
  Adam P. Showman\altaffilmark{10},\\
  Gregory S. Tucker\altaffilmark{1}}


\affil{1.  Department of Physics, Box 1843, Brown University, Providence, RI 02904, USA}
\affil{2.  Space Telescope Science Institute, Baltimore, MD 21218, USA}
\affil{3. Jet Propulsion Laboratory, California Institute of Technology, 4800 Oak Grove Drive, Pasadena, CA 91109, USA}
\affil{4. Department of Astronomy and Carl Sagan Institute, Cornell University, 122 Sciences Drive, Ithaca, NY, 14853, USA}
\affil{5. Center of Excellence in Information Systems, Tennessee State University, 3500 John A. Merritt Blvd., P.O. Box 9501, Nashville, TN 37209, USA}
\affil{6.  Department of Physics, Department of Earth \& Planetary Sciences, McGill University, 3450 rue University, Montreal, QC CAN}
\affil{7. Dept. of Earth, Atmospheric and Planetary Sciences, Massachusetts Institute of Technology, 77 Massachusetts Ave., Cambridge, MA, 02139.}
\affil{8.  Department of Astronomy and Astrophysics, University of California Santa Cruz}
\affil{9. California Institute of Technology, Pasadena, CA, USA}
\affil{10.  Department of Planetary Sciences and Lunar and Planetary Laboratory, University of Arizona, AZ 85721}

\affil{*  NASA Earth and Space Science Fellow}

\begin{abstract}

Here we present the analysis of multi-epoch secondary eclipse observations of HD 189733b and HD 209458b as a probe of temporal variability in the planetary climate using both \Spitzer channels 1 and 2 (3.6 and 4.5 \micron). We expect hot Jupiter atmospheres to be dynamic environments exhibiting time varying weather. However, it is uncertain to what extent temporal variability will be observable when considering disc integrated observations. We do not detect statistically significant variability
and are able to place useful upper limits on the IR variability amplitudes in these
atmospheres.
There are very few planets with multi-epoch observations at the required precision to probe variability in dayside emission. The observations considered in this study span several years,  providing insight into temporal variability at multiple timescales.  
In the case of HD 189733b, the best fit eclipse depths for the channel 2 observations exhibit a scatter of  102 ppm about a median depth of 1827 ppm and in channel 1 exhibit a scatter of  88 ppm about a median depth of 1481 ppm.  For HD 209458b, the best fit eclipse depths for the channel 2 observations exhibit a scatter of  22 ppm about a median depth of 1406 ppm and in channel 1 exhibit a scatter of  131 ppm about a median depth of 1092 ppm.  The precision and scatter in these observations allow us to constrain variability to less than (5.6\% and 6.0\%) and  (12\% and 1.6\%) for channels (1,2) of HD 189733b and HD 209458b respectively.  

\end{abstract}

 \maketitle

\section{Introduction}\label{sec: Intro}

Studying exoplanet atmospheres is challenging since we often are not able to spatially resolve them.  They are typically too faint to disentangle their light from that of their much brighter host star.  Transiting exoplanets provide a unique opportunity to study exoplanet atmospheres in spite of this challenge.  Disc integrated secondary eclipse observations provide valuable information on temperature, albedo, and chemical composition averaged over the entire hemisphere.  Primary transits give a limb averaged atmospheric molecular spectrum and pressure - temperature profile.  However, we must employ other novel techniques to begin to probe physical and chemical processes and structures at smaller scales.  Temporal variability is one way to probe weather features and their their movements without spatially resolving them.  Variability is also a tool to study portions of the atmosphere inaccessible via other observational techniques.  Transmission spectroscopy observations in the near infrared typically probe millibar pressure levels high in the atmosphere.  Processes deeper in the atmosphere are obscured due to the long path lengths of the transit geometry.  Time varying changes in emission may be indicative of processes occurring deeper in the atmosphere that are, otherwise, unobservable with current methods and insturments.

Close-in giant planets (hot Jupiters) are interesting objects with which to study atmospheric dynamics. They are assumed to be tidally locked based on their short orbital periods (<5 days) and minimal separation from their host stars. As a result, they experience a constant radiative forcing on their permanent dayside.  The combination of the rotation rate and radiative forcing of these planets are predicted to produce large scale weather structures unlike anything in our solar system.

 There have been many efforts to model the circulation and temperature structure of hot Jupiters with many of these efforts based on the properties of the two most well studied targets, HD 189733b and HD 209458b (HD 189733b\& HD 209). Simulations by \cite{2002Show, 2003Cho} and \cite{2003Menou} predicted that the Rossby deformation radius and Rhines scale of hot Jupiters should be comparable to the planetary radius resulting in atmospheric dynamics comprised of a few jets and large scale polar vortices.  They predicted that vortices would be large enough that their migration could have an affect on the observed eclipse depth.   \citet{2007RauscherVar}  demonstrated that, based on those model predictions, changes in eclipse depth as great as 20\% could be observed.  Later simulations by \cite{Showman2009} predicted far more stable atmospheric structures.  Variability was predicted to be less than several percent for HD 189733b at 8 $\mu$m and the result of wave dynamics deep in the atmosphere rather than migrating vortices.

\cite{Agol2010} observationally probed model predictions with a multi - epoch set of transit and eclipse observations of HD 189733b at 8 \micron.  The results of that work placed an upper limit on eclipse depth variability at 2.7\%, effectively ruling out larger predictions.  
There have been few other attempts to study temporal variability due to the lack of multiple observations over a sufficient temporal baseline.  

We choose as our targets for this study HD 189733b and HD 209458b both because of their suitability for observations with the \Spitzer observatory and because they are the only targets with a large number of observations with a single instrument spanning a number of years.  These two planets are the most well studied and characterized exoplanets to date.  Both orbit bright stars, K$_{\rm{mag}}$ of 5.5 and 6.3 for HD 189733b and HD 209458b respectively, making them ideal targets for characterization and have previously observed secondary eclipse depths >1000 ppm in both warm \Spitzer channels 1 and 2.  

HD 189733b is arguably the most thoroughly studied exoplanet.  It  has been observed with photometric \citep[e.g.][]{2007ApJ...668L.179E,2009ApJ...699..478D, 2011A&A...526A..12D} and spectroscopic \citep[e.g.][]{2012Gibson} transits as well as secondary eclipse \citep[e.g.][]{2008ApJ...686.1341C, Agol2010} and phase curve \citep{Knutson2007, 2009Knutson, Knutson2012} observations across a multitude of wavelengths.  In addition to full orbit phase curves at multiple wavelengths, it is the only planet to have been mapped via the eclipse mapping technique \citep{2012Majeau, 2012Dewit, 2018Rauscher}.
HD 209458b was the first transiting exoplanet discovered \citep{2000Charb}. It has also been observed in both transit and eclipse geometries photometrically \citep[e.g.][]{2008Charb, 2014DL} and spectroscopically \citep[e.g.][]{2013Dem209, 2016Line} , as well as full orbit phase curve observations \citep{2014Zellem}.  

The large amount of observational and theoretical effort invested into these two canonical hot Jupiters make them perfect choices for a study of this nature.  There are eclipse observations from multiple programs spanning years from which to build the necessary temporal baseline.  Additionally, their orbital properties are well studied and well constrained allowing for the detection of small signals in their lightcurves due to planetary thermal structure that can be clearly distinguished from other potential sources such as orbital eccentricity.   

Understanding variability will be essential in current and future attempts to leverage multiple epoch observations to build precision for high resolution techniques. 
 When stacking observations, whether it be transmission spectroscopy or emission photometry, one makes the assumption that each observation is an independent measurement of a constant signal.  Showing that any temporal variability is below the level of precision of your observations allows them to be combined in analysis.  If this set of \Spitzer observations are not sensitive to temporal variability they can be combined to achieve the precision necessary to perform eclipse mapping.

Eclipse mapping \citep{2012Majeau, 2012Dewit} uses the deviations to the shape of ingress and egress caused by a non uniform dayside temperature distribution to create two dimensional thermal maps.  In general, one needs $\sim$10 points over the ingress/egress with a precision of at least a tenth of the eclipse depth to achieve the necessary resolution to map hot Jupiters.  The only way to achieve that precision with current observatories is to stack multiple observations. 
In the {\sl James Webb Space Telescope (JWST)} era, observing time will be expensive and full orbit phase curves may prove too costly.  However, mapping may be done much more efficiently with as little as two eclipse observations.  Understanding temporal variability then is crucial in knowing if stacking multi-epoch observations is justifiable in that context. The observations considered here give the best current insight into orbit to orbit changes in hot Jupiter atmospheres.

\section{Observations}\label{sec: Obs}
Here we consider all existing secondary eclipse observations of HD 189733b and HD 209458b in channels 1 and 2 (3.6 or 4.5$\, \mu{\rm m}$ bandpass) .
The observations analyzed here are part of Programs  60021 (PI: H. Knutson), 10103 (PI: N. Lewis), 90186 (PI:  K. Todorov), and 70100 (PI: M. Swain). 
The details of each Astronomical Observing Request (AOR) are displayed in Table \ref{obstab209} and \ref{obstab189}.  All of the observations were carried out in sub-array mode ($32\times 32$ pixels, $39{\tt "} \times 39{\tt "}$).  Observations in Program 10103 utilize a 30 minute peak-up observation preceding them to stabilize the image on the detector `sweet spot' and decreases the likelihood of a ramp in the data \citep{Ingalls}.  Program 60021 observations are full orbit phase curves from which we extract the AORs containing the eclipse from the larger data set.  Both HD 189733 and HD 209458 are bright targets with K$_{mag}$ of 5.5 and 6.3 respectively.  As a result, frame times for all of the observations are 0.1 seconds.  The two exceptions are the observations of HD 209458b in channel 2 that were conducted with a 0.4 second frame time and the observations of HD 189733b from program 70100 that were conducted such that the stellar centroid was near the corner of the pixel so that a longer (0.4 s) exposure time could be used.  This technique proved problematic in that the systematics were not easily corrected with existing techniques and, as a result, several of these observations were discarded from this study.

\section{Data Analysis Methods}\label{sec: Analysis}

\subsection{Photometric Extraction}

For each AOR we began with Basic Calibrated Data (BCD) available on the Spitzer Heritage Archive.  Each BCD file contains a cube of 64 frames of $64\times 64$ pixels.  Each set of 64 images comes as a single FITS file with a time stamp corresponding to the start of the first image.  We determine the time of each frame in the set by adding the appropriate multiple of the frame time to the time stamp of the first image.
Each frame was corrected for bad pixels or NaN values by masking the invalid pixels.  Each frame is background subtracted based on the median pixel value outside of a 10 x 10 pixel box around the stellar centroid.  Pixel values in the background region that deviate from the mean by more than 5 $\sigma$ are clipped and the median value remaining is taken to be the background. Stellar centroid positions for each frame are determined by a first moment center of mass calculation:  
\begin{equation}
x_{\rm cen}=\frac{\sum_{j,k}(I_{jk}j)}{\sum_{j,k}I_{jk}}; \qquad
y_{\rm cen}=\frac{\sum_{j,k}(I_{jk}k)}{\sum_{j,k}I_{jk}}.
\end{equation}
We perform two-dimensional Gaussian Centroiding as well but find the first moment calculation to be more stable and provides less scatter in the extracted photometry.  

\capstartfalse
\begin{deluxetable}{cccc}[!h]
\tablecaption{Summary of Observations\\HD 209458b \label{obstab209}}
\tablehead{
\colhead{Program} & \colhead{AOR} & \colhead{Channel} & \colhead{Date}} 
60021			& 41629440	   & 1				  &	1/12/11 	  \\ 
60021			& 41628416	   & 1				  &	1/15/11 	  \\ 
90186			& 48013824	   & 1 				  &	8/28/13 	  \\ 
10103           & 50496512     & 1                & 1/19/14       \\ 
10103           & 50496256     & 1                & 2/27/14       \\ 
10103           & 50496000     & 1                & 2/6/14        \\ 
10103           & 50495744     & 1                & 8/22/14       \\ 
10103           & 50495488     & 1                & 8/29/14       \\ 
10103           & 50494976     & 1                & 9/5/14        \\ 
10103           & 50494208     & 1                & 9/27/14       \\ 
10103           & 50493440     & 1                & 1/24/15       \\ 
10103          & 50492928     & 1                & 1/28/15       \\ 
10103           & 50490880     & 2                & 1/26/14       \\ 
10103           & 50490624     & 2                & 2/2/14        \\ 
10103           & 50490368     & 2                & 8/19/14       \\ 
10103           & 50490112     & 2                & 8/26/14       \\ 
10103           & 50489856     & 2                & 9/12/14       \\ 
10103           & 50489088     & 2                & 1/7/14        \\ 
90186           & 48014336     & 2                & 8/31/13       \\ 
60021			& 38703616	   & 2				  & 1/21/10 	  

\enddata
\end{deluxetable}
\capstarttrue
\capstartfalse
\begin{deluxetable}{cccc}[!h]
\tablecaption{Summary of Observations\\HD 189733b \label{obstab189}}
\tablehead{
\colhead{Program} & \colhead{AOR} & \colhead{Channel} & \colhead{Date}} 
10103 &	50495232 & 1 &	1/7/14 \\
10103 &	50494464 &	1 & 1/14/14 	 \\
10103 &	50493696 &	1 & 1/25/14 	\\
10103 &	50493184 &	1 & 7/15/14 	 \\
60021 &	41592320 &	1 & 12/28/10 	\\
60021 &	41591296 &	1 & 12/30/10  \\
70100 &	40150528 &	1 & 11/24/10  \\
70100 &	40151040 &	1 & 1/3/11  \\
70100 &	40151296 &	1 & 6/23/11  \\
70100 &	40152064 &	1 & 6/27/11  \\
10103 &	50489344 &	2 & 1/12/14  \\
10103 &	50488832 &	2 & 1/18/14  \\
10103 &	50488576 &	2 & 7/13/14  \\
10103 &	50488320 &	2 & 7/26/14  \\
10103 &	50488064 &	2 & 8/11/14  \\
60021 &	38390784 &	2 & 12/22/09  \\
60021 &	38390016 &	2 & 12/24/09  

\enddata
\end{deluxetable}
\capstarttrue

The noise pixel parameter ($\tilde{\beta}$) \citep{Lewis2013, KilSpitzer} is calculated for each frame as 
\begin{equation}\tilde{\beta}=\frac{\left(\sum P_i\right)^2}{\sum \left(P_i^2\right)}, \end{equation} 
where each $P_i$ is the response measured in each pixel across the frame. We also save the 5 x 5 array of background subtracted pixel values about the stellar centroid for each frame.  Each array is normalized such that they sum to one.  We then perform aperture photometry about the stellar centroid using the aperture\_photometry function from the Astropy package Photutils. Circular apertures of fixed radii ranging from 1.8 - 2.8 pixels and variable radii apertures ranging from  $\beta_{\rm pix} \times$ \{0.7...1.2\} and  $\beta_{\rm pix} +$ \{-0.6...1.4\}.
The resultant time series photometry is then filtered by removing any points that deviate by more than 3 $\sigma$ from the median values of flux or x,y position.  

\subsection{Systematics Correction and Model Fitting}

The intrapixel sensitivity variation \citep{2012Ingalls}, the change in measured flux as a function of stellar centroid position and methods of correction, are well documented \citep[e.g.][]{2016Ingalls, KilSpitzer}.  Here, we employ two independent methods of correction for this systematic variation;  the Nearest Neighbors method (NNBR), otherwise known as Gaussian Kernel Regression with data \citep{Lewis2013} and Pixel Level Decorrelation (PLD) \citep{Deming}. 

Each eclipse fit was based on the model of \cite{Mandel} for a uniform occultation implemented in python by the BATMAN package \citep{2015Kreidberg}.  The orbital, stellar, and planetary parameters listed in table \ref{ephem} were used as input fixed parameters to the model.  \Spitzer IRAC data is known to have an exponential ramp in flux over the first 30--60  minutes of observing; however the peak-up technique has alleviated this problem to some extent \citep{KilSpitzer, Ingalls}.  As a precaution, each data set was fit in several ways: without alteration, trimmed at 20 minutes from the beginning of the observation, and with an exponential ramp model of the form \begin{equation}
R_{model}=1-a_{1}\times e^{(-\frac{t}{a_2})}.
\end{equation} 

PLD has a quadratic visit long temporal variation included in all cases rather than the exponential ramp and combines the systematics and astrophysical models by a linear expression
\begin{equation} \label{seven}
\Delta S^{t} = \sum \limits _{i=1}^n c_i\hat{P}^t_i +DE(t) +ft+gt^2+h,
\end{equation} where the $\hat{P}^t_i$'s are the response of the $i^{th}$ pixel in a 3 $\times$ 3 grid around the centroid of the image and $DE(t)$ is the eclipse model and $\Delta S^t$ is the total flux, including the astrophysical signal and all systematics, at each time $t$.  

For each AOR the eclipse model and systematics model were combined and best fit values for all free parameters were determined using a non-linear, least squares fitting algorithm.  The standard deviation of the normalized residuals (SDNR) was used as a metric for selecting the best fit out of the 30 different apertures for each AOR.  The Bayesian Information Criterion (BIC) was used to determine the appropriate treatment between various trimming and ramp model options. We find that no trimming or exponential ramp are necessary in all cases of Channel 2 observations and the quadratic ramp with PLD is favored in all Channel 1 observations.  The results from the best fit aperture were passed to a Markov Chain Monte Carlo implemented by emcee \citep{emcee} to derive uncertainties of each free parameter.  We use a number of walkers at least twice the number of free parameters and run for $10^5$ steps per walker before testing for convergence using Gelman Rubin statistics with a threshold for acceptance of 1.01 \citep{gelman1992}.

\capstartfalse
\def\arraystretch{1.5}
\tabletypesize{\footnotesize}
\begin{deluxetable}{lcc}[]
\tablecaption{Ephemerides\label{ephem}}
\tablehead{
\colhead{Parameter} & \colhead{HD 189733b} & \colhead{HD 209458b} } \label{ephem}
\startdata
$T_{\star}$(K) & 5040$\pm$50 & 6065$\pm$50\\
$M_\star$ (M$_{\rm Sun})$ & 0.806$\pm$0.048 & 1.131$\pm$0.026 \\
$R_\star$ (R$_{\rm Sun})$ &0.756$\pm$0.018  & 1.155$\pm$0.015 \\
$M_p$ (M$_{\rm Jup})$ &1.144$\pm$0.056  & 0.690$\pm$0.024 \\
$R_p$ (R$_{\rm Jup})$ &1.138$\pm$0.027  & 1.359$\pm$0.015 \\
$R_p/R_{\star}$ &0.024122$\pm$5.8$\times$10$^{-5}$  & 0.014607$\pm$2.4$\times$10$^{-5}$ \\
log(g) (log(c/s$^2$))&3.339$\pm$0.03  & 2.969$\pm$0.0187\\
Period (days)& 2.21857567$\pm$1.5$\times$10$^{-7}$& 3.52474859$\pm$ 3.8$\times$10$^{-7}$\\
$i$ ($^{\circ}$) & 85.7100$\pm$0.0023 & 86.710$\pm$ 0.05 \\
m$\sin{i}$ (M$_{\rm Jup})$&1.140 $\pm$0.056  & 0.689$\pm$0.024 \\
$a/R_{\star}$  &8.84$\pm$0.27 & 8.81$\pm$0.186  \\
\enddata
\end{deluxetable}
\capstarttrue

\section{Results}\label{sec: Res}

\begin{figure}[!t]
\vspace{0.1in}
 \centering
\includegraphics[width=0.45\textwidth]{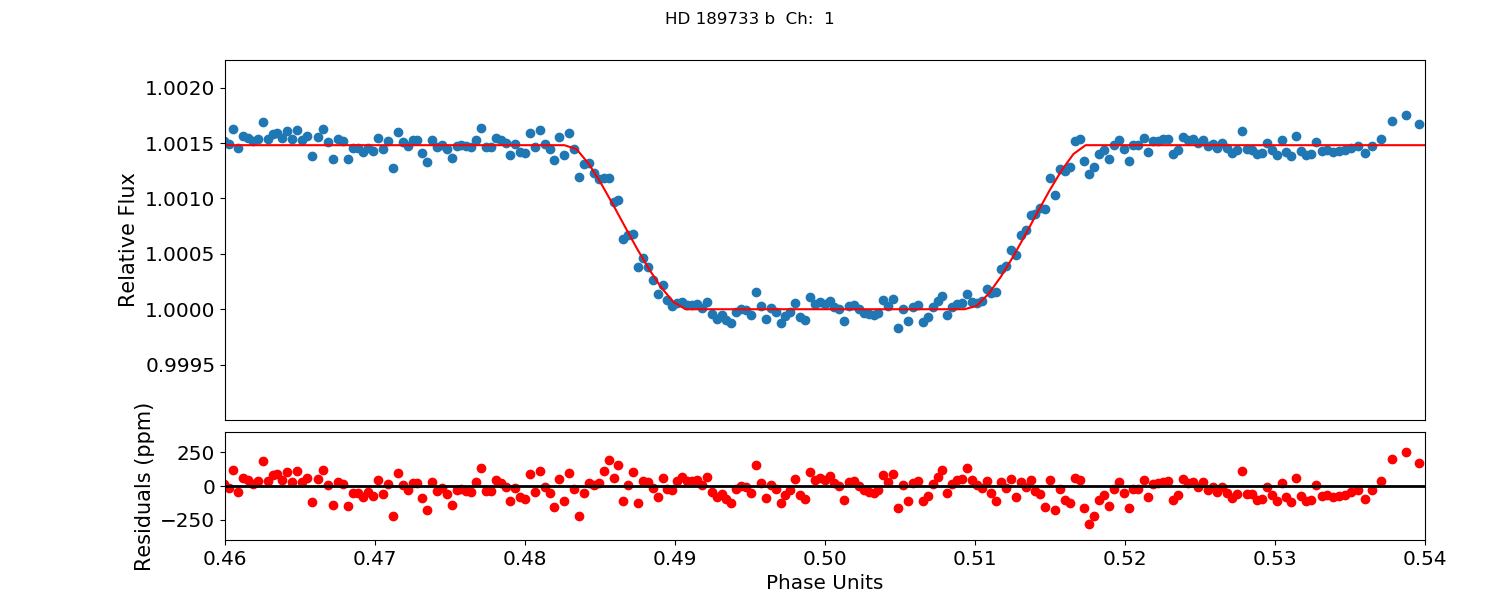}
\includegraphics[width=0.45\textwidth]{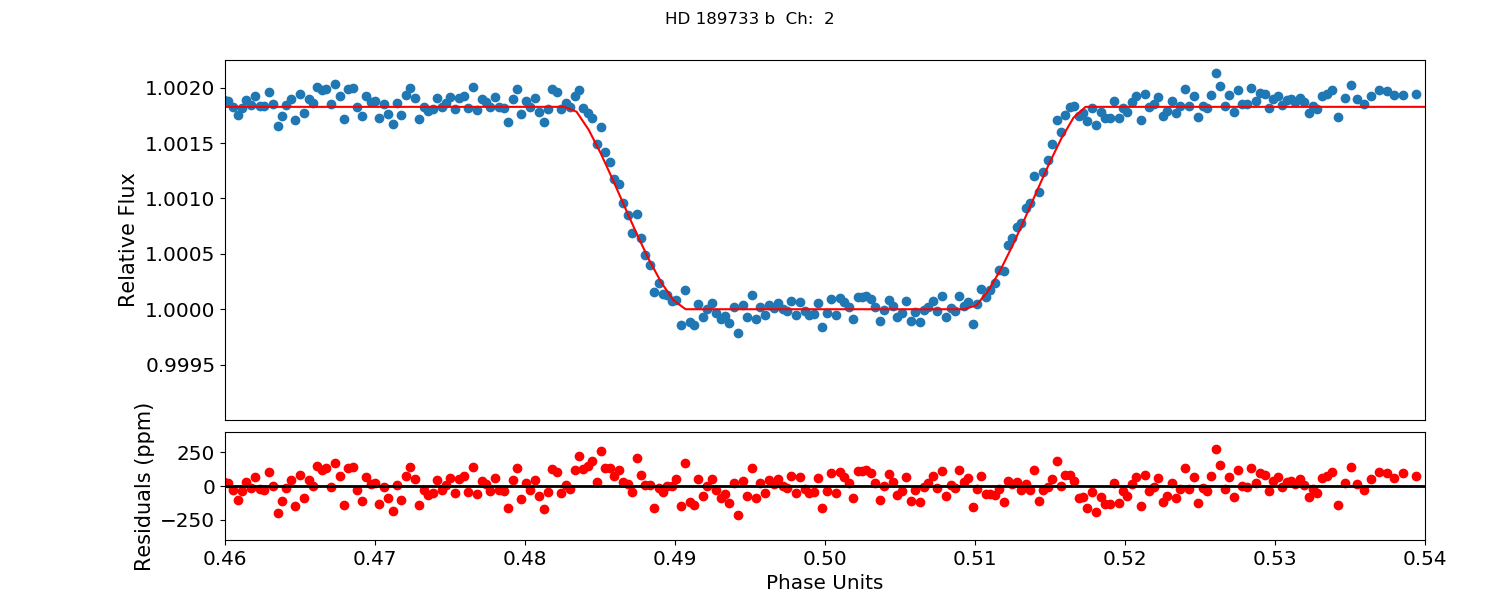}
\includegraphics[width=0.45\textwidth]{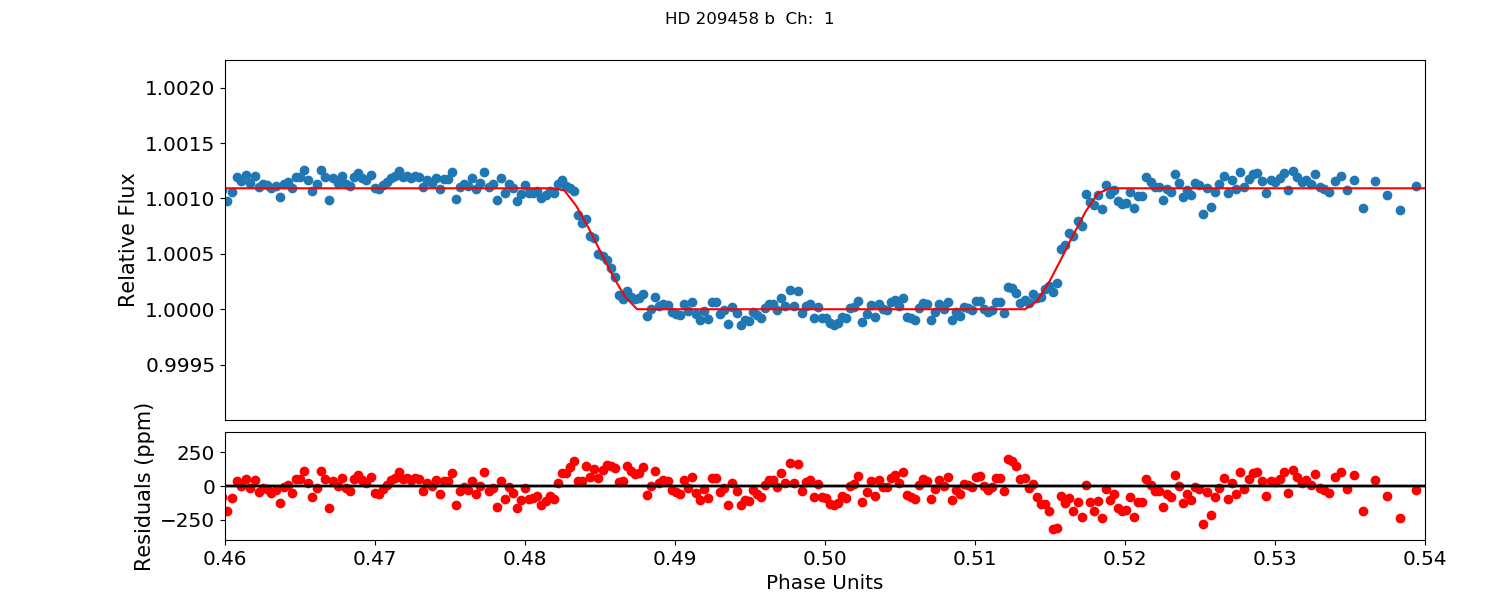}
\includegraphics[width=0.45\textwidth]{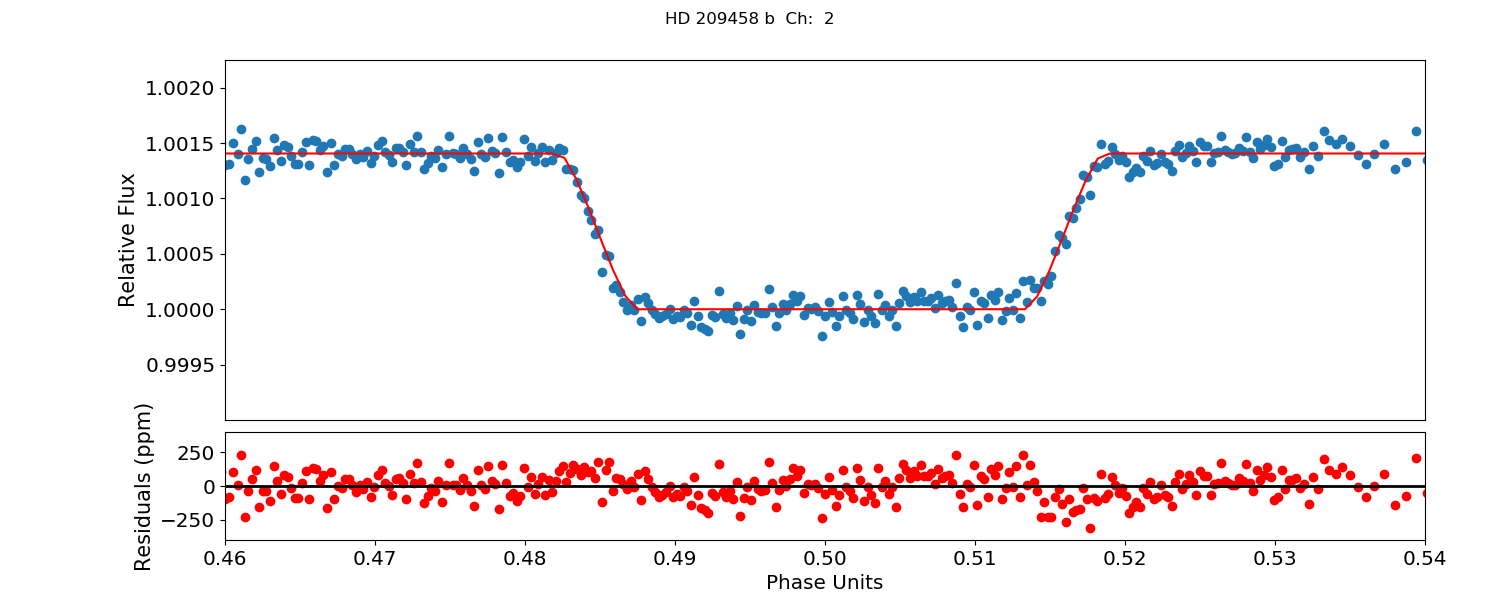}
\caption{Each of the above panels shows the lightcurve after all data points for each planet/channel are binned to $\sim$1 minute bins with systematics removed.  The best fit model based on the weighted averages as stated in Section \ref{sec: Res} is plotted in red.  The residuals from each individual fit are combined and binned at the same interval.}
 \label{best_fits}
 \vspace{0.1in}
 \end{figure}

Here we present the best fit eclipse depths and center of eclipse times, along with their corresponding uncertainty, for each of the observations. 
The results presented in Table \ref{209res2} are derived using the PLD method with a quadratic temporal term.  No additional trimming or ramps are modeled.  The best fit aperture was consistently a fixed radius aperture between 2.2 and 2.4 pixels. We include in Table \ref{209res2} the standard deviation of the normalized residuals (SDNR) and the $\beta_{\rm red}$ factor, as defined in \citet{gillon2010}, as a measure of correlated noise remaining in the data after systematic corrections.  Included in Figure \ref{best_fits} are representative fits for each of the two targets in each of the two channels with the systematics removed and binned at two minute intervals. In the case of HD 189733b, the best fit eclipse depths for the channel 2 observations exhibit a scatter of 102 ppm about a median depth of 1827 ppm and in channel 1 exhibit a scatter of 88 ppm about a median depth of 1481 ppm.  For HD 209458b, the best fit eclipse depths for the channel 2 observations exhibit a scatter of  22 ppm about a median depth of 1406 ppm and in channel 1 exhibit a scatter of  131 ppm about a median depth of 1092 ppm.

\capstartfalse
\def\arraystretch{1.2}
\tabletypesize{\footnotesize}
\begin{deluxetable}{lcccc}[]
\tablecaption{Results of Best Fit Eclipse Depth and Time}
\tablehead{
\colhead{AOR} & \colhead{Eclipse Depth} & \colhead{Eclipse Time} & \colhead{SDNR} & \colhead{$\beta_{\rm red}$} \\ \colhead{} & \colhead{(ppm)} & \colhead{ O-C (min)} & \colhead{} & \colhead{}} 
\startdata
\multicolumn{5}{c}{HD 209458b Ch 2}\\
\hline
38703616 & 1370 $\pm$ 37 & 2.22 $\pm$ 1.04 & 0.003240 & 1.43\\
48014336 & 1401 $\pm$ 42 & 1.44 $\pm$ 0.83 & 0.003085 & 1.17\\
50489088 & 1424 $\pm$ 60 & 3.03 $\pm$ 0.90 &  0.003085 & 1.16\\
50490880 & 1380 $\pm$ 58 & 1.62 $\pm$ 0.79 & 0.003053 & 1.27\\
50490624 & 1443 $\pm$ 56 & 3.14 $\pm$ 0.82  & 0.003060 & 1.27\\
50490368 & 1417 $\pm$ 45 & 2.60 $\pm$ 0.96 & 0.003151 & 1.05\\
50490112 & 1420 $\pm$ 52 & 2.81 $\pm$ 0.72 & 0.003125 & 1.18\\
50489856 & 1424 $\pm$ 47 & 1.57 $\pm$ 0.65 & 0.003116 & 1.09\\
\hline
\multicolumn{5}{c}{HD 209458b Ch 1}\\
\hline
48013824 & 1050 $\pm$ 73 & 4.37 $\pm$ 1.61 & 0.004995 & 2.59\\
50496512 & 909 $\pm$ 60 & 1.48 $\pm$ 1.27 & 0.004873 & 1.99\\
50496000 & 1162 $\pm$ 77 & 3.38 $\pm$ 1.39 &  0.004834 & 1.96\\
50496256 & 1209 $\pm$ 67 & 2.81 $\pm$ 1.13 & 0.004805 & 1.95\\
50495744 & 845 $\pm$ 92 & 4.21 $\pm$ 1.84  & 0.005055 & 2.29\\
50495488 & 1008 $\pm$ 76 & 1.81 $\pm$ 1.37 & 0.004946 & 1.78\\
50494976 & 1049 $\pm$ 78 & 2.48 $\pm$ 1.30 & 0.004898 & 2.03\\
50494208 & 1209 $\pm$ 72 & 1.61 $\pm$ 1.36 & 0.0049606 & 2.03\\
50493440 & 1301 $\pm$ 76 & 2.20 $\pm$ 1.15 & 0.004983 & 1.87\\
50492928 & 994 $\pm$ 105 & 6.82 $\pm$ 2.61 & 0.004939 & 2.74\\
41629440 & 1216 $\pm$ 77 & 1.28 $\pm$ 2.32 & 0.006171 & 4.14\\
41628416 & 1128 $\pm$ 72 & 1.99 $\pm$ 2.72 & 0.006168 & 3.68\\
\hline
\multicolumn{5}{c}{HD 189733b Ch 2}\\
\hline
38390784 & 1812 $\pm$ 50 & 1.52 $\pm$ 0.69 & 0.004596 & 1.65\\
38390016 & 1806 $\pm$ 64 & 0.31 $\pm$ 0.97 & 0.004698 & 1.74\\
50489344 & 1716 $\pm$ 55 & 0.96 $\pm$ 0.45 &  0.004631 & 1.06\\
50488832 & 1815 $\pm$ 61 & -0.37 $\pm$ 0.61 & 0.004606 & 1.41\\
50488576 & 1984 $\pm$ 75 & 0.31 $\pm$ 0.64  & 0.004630 & 1.31\\
50488320 & 1774 $\pm$ 50 & 0.24 $\pm$ 0.55 & 0.004572 & 1.12\\
50488064 & 2016 $\pm$ 65 & -0.02 $\pm$ 0.48 & 0.004654 & 1.00\\
\hline
\multicolumn{5}{c}{HD 189733b Ch 1}\\
\hline
41592320 & 1574 $\pm$ 305 & 1.40 $\pm$ 6.0 & 0.004228 & 4.55\\
41591296 & 1431 $\pm$ 86 & 0.29 $\pm$ 1.74 & 0.003382 & 2.13\\
40152064 & 1271 $\pm$ 310 & 0.16 $\pm$ 4.08 &  0.001708 & 3.76\\
50495232 & 1514 $\pm$ 105 & -0.37 $\pm$ 1.35 & 0.003370 & 1.51\\
50494464 & 1473 $\pm$ 73 & 0.81 $\pm$ 0.93  & 0.003298 & 1.50\\
50493696 & 1498 $\pm$ 53 & 0.80 $\pm$ 0.67 & 0.003266 & 1.54\\
50493184 & 1491 $\pm$ 115 & -0.74 $\pm$ 1.37 & 0.003306 & 1.85

\enddata
\label{209res2}
\end{deluxetable}
\capstarttrue

\begin{figure}[!t]
 \centering
\includegraphics[trim=0.5in 0.0in 0.5in 0.0in,clip, width=0.47\textwidth]{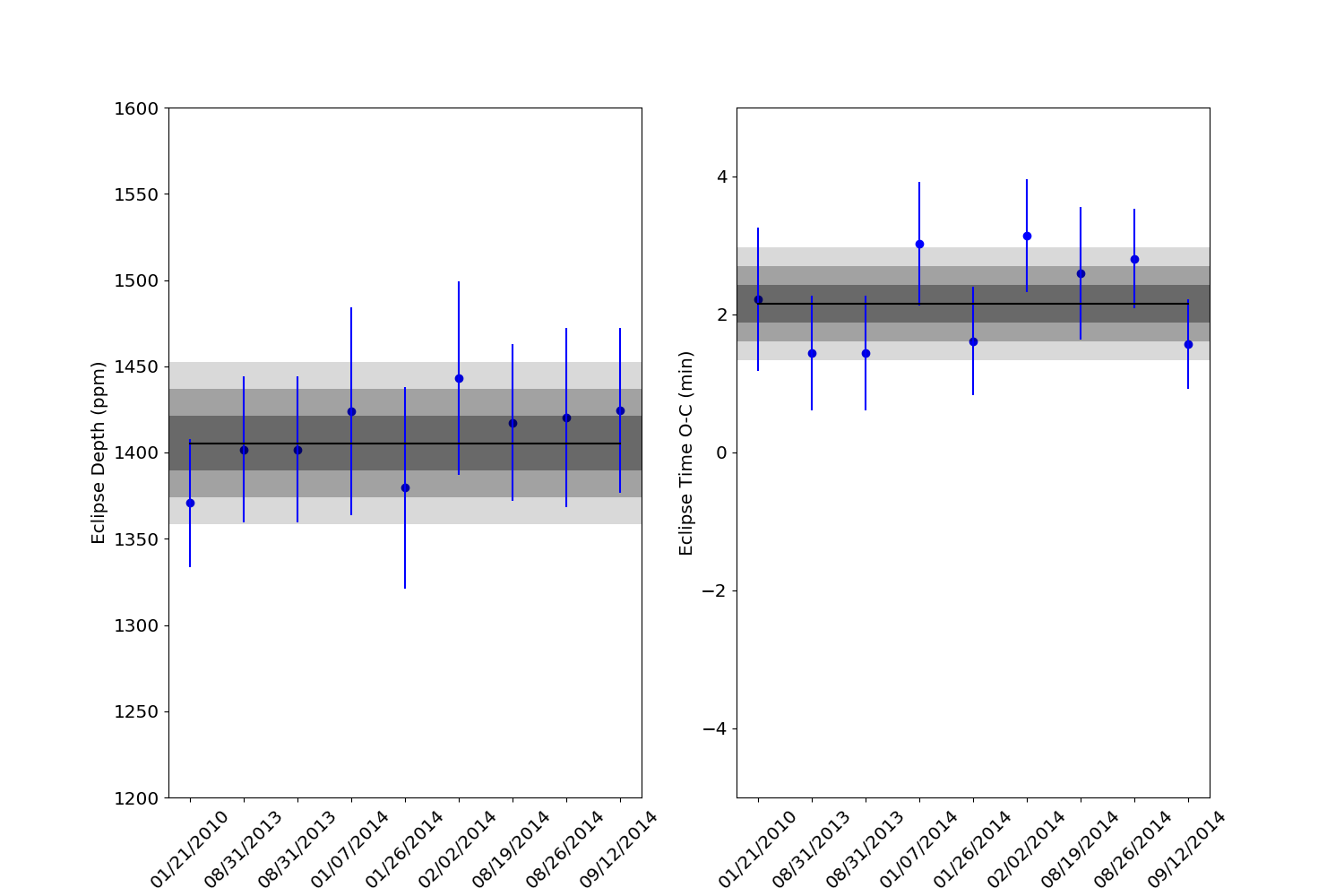}
\includegraphics[trim=0.5in 0.0in 0.50in 0.55in,clip, width=0.47\textwidth]{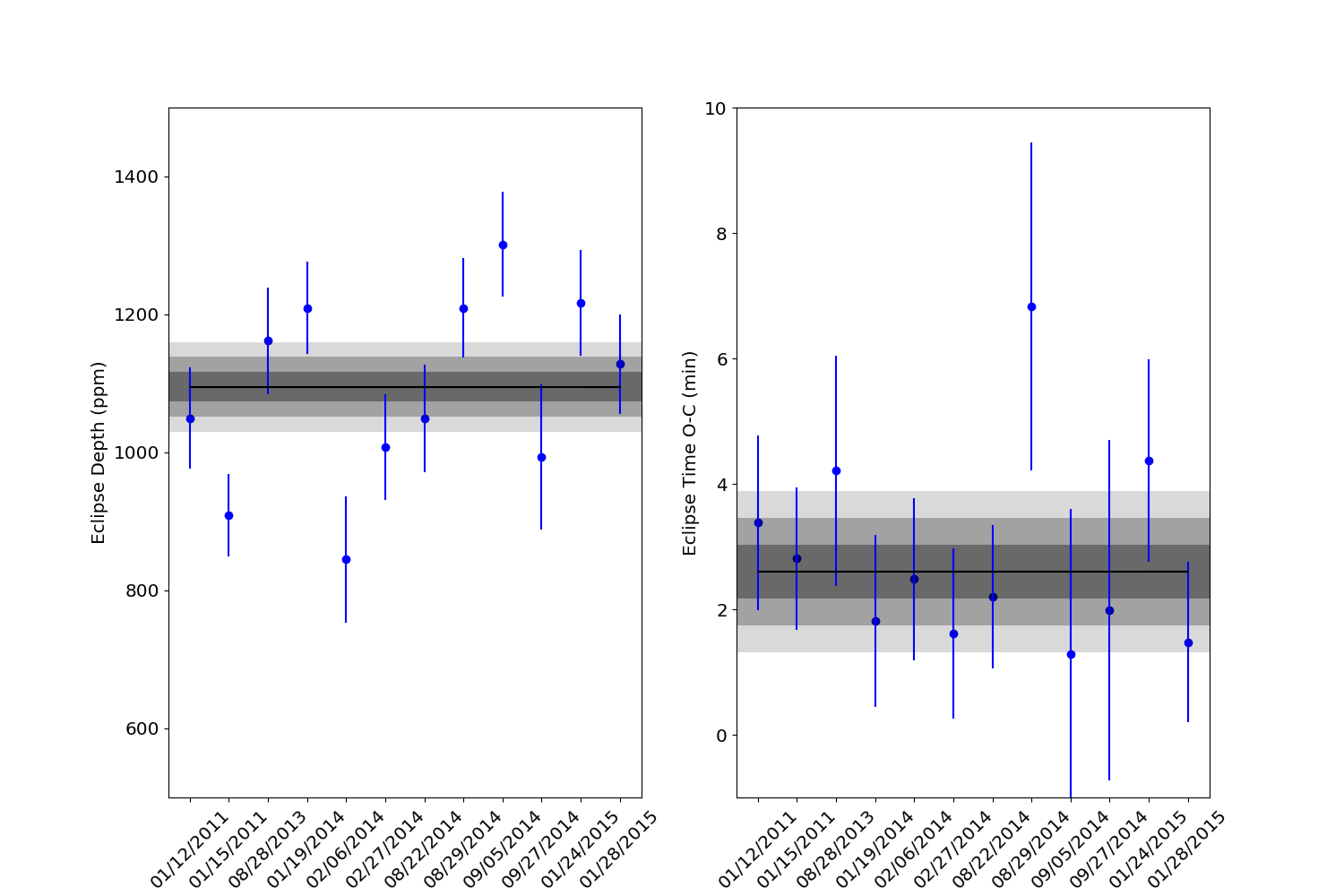}
\caption{Eclipse Depths and timing for the 9 Ch 2 Eclipses (top) and the the 12 Ch 1 Eclipses (bottom) of HD 209458b with multiple methodologies.  The  shaded gray areas represent the 1,2 and 3 sigma areas of a distribution derived from the mean and uncertainty of the measurements as a whole.  }
 \label{figure:alldepths1}
 \end{figure}

\begin{figure}[!t]
 \centering
\includegraphics[trim=0.5in 0.0in 0.5in 0.0in,clip, width=0.48\textwidth,width=0.45\textwidth]{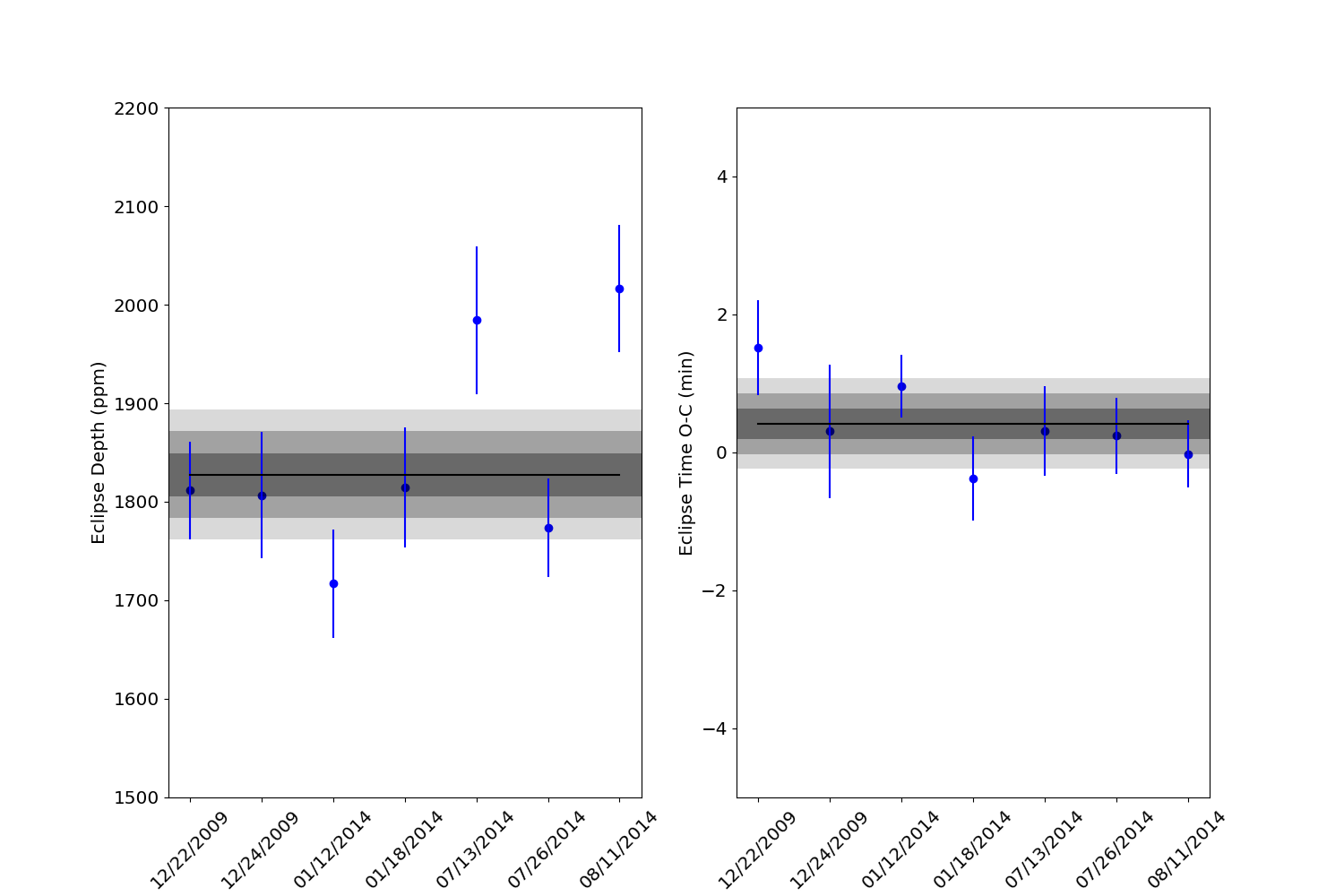}
\includegraphics[trim=0.5in 0.0in 0.5in 0.5in,clip, width=0.47\textwidth]{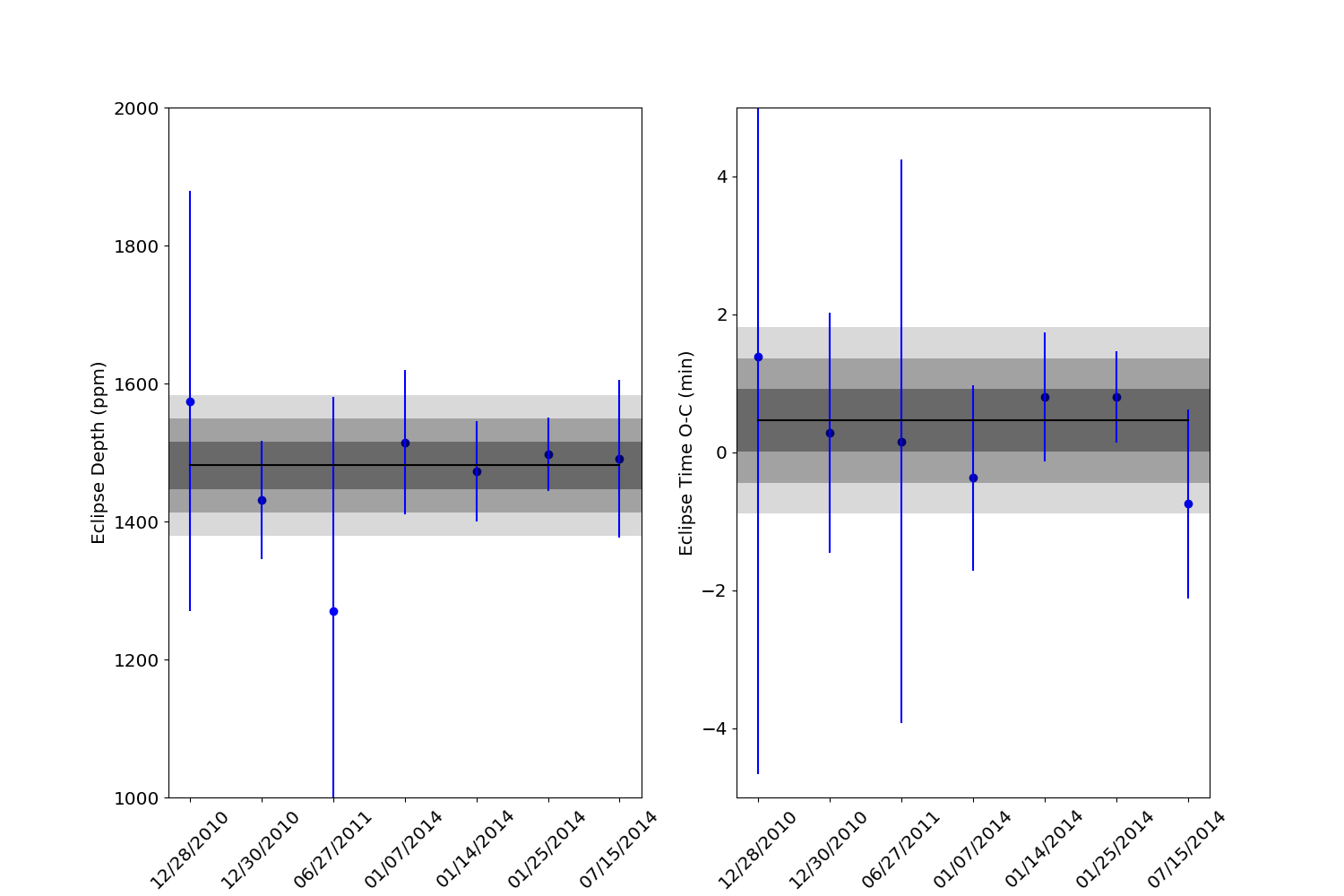}
\caption{Eclipse Depths and timing for the 7 Ch 2 Eclipses (top) and the the 7 Ch 1 Eclipses (bottom) of HD 189733b with multiple methodologies.  The  shaded gray areas represent the 1,2 and 3 sigma areas of a distribution derived from the mean and uncertainty of the measurements as a whole.  }
 \label{figure:alldepths2}
 \end{figure}

As shown in Figures \ref{figure:alldepths1} and \ref{figure:alldepths2}, the scatter in the data is consistent with the uncertainty in each measurement in all cases except for the channel 1 HD209 data. Given the number of observations at the achieved precision, we would expect the resultant eclipse depths to represent a sampling of a distribution represented by the shaded areas in Figures \ref{figure:alldepths1} and \ref{figure:alldepths2}.  The channel 1
observations of HD209, which exhibit more scatter than indicated by the error bars, is thought to be due to unresolved systematics. This is further supported by the larger values of $\beta_{\rm red}$ in channel 1 observations in comparison to channel 2.

\subsection{Constraints on Temporal Variability}

\begin{figure}[h!]
 \centering
\includegraphics[trim=0.0in 0.0in 0.0in 0.0in,clip, width=0.45\textwidth]{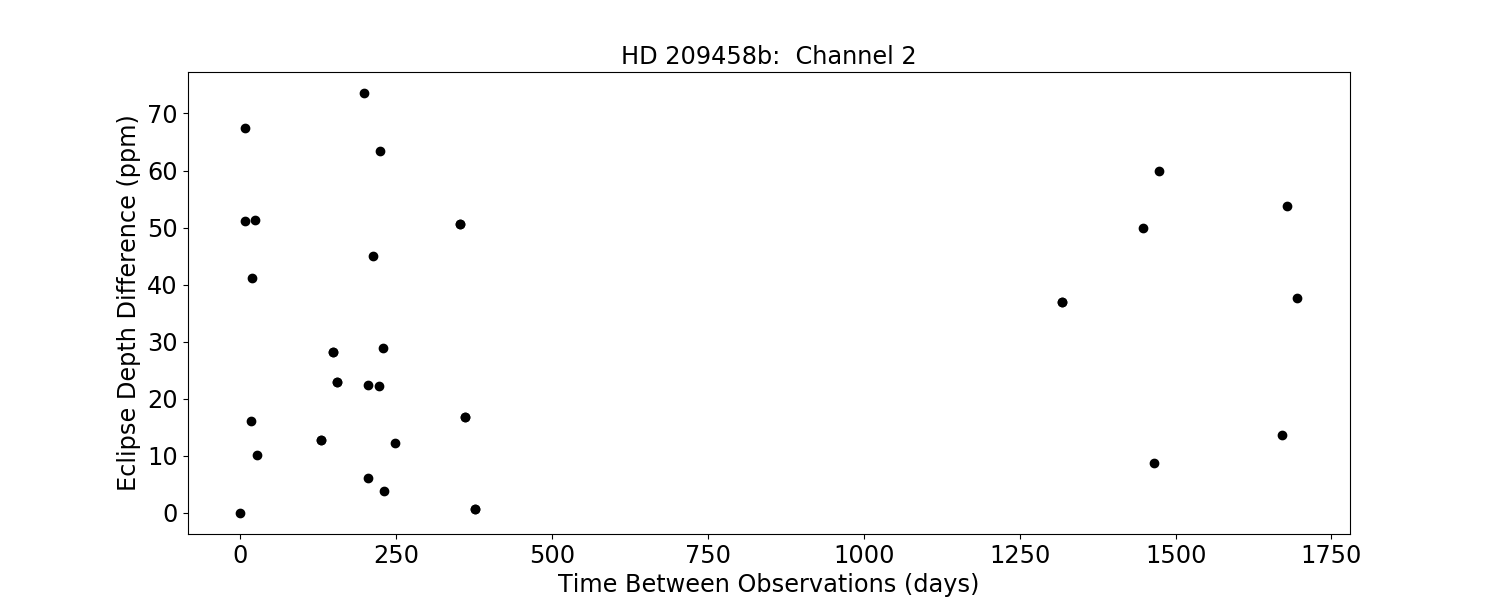}\\
\includegraphics[trim=0.0in 0.0in 0.0in 0.0in,clip, width=0.45\textwidth]{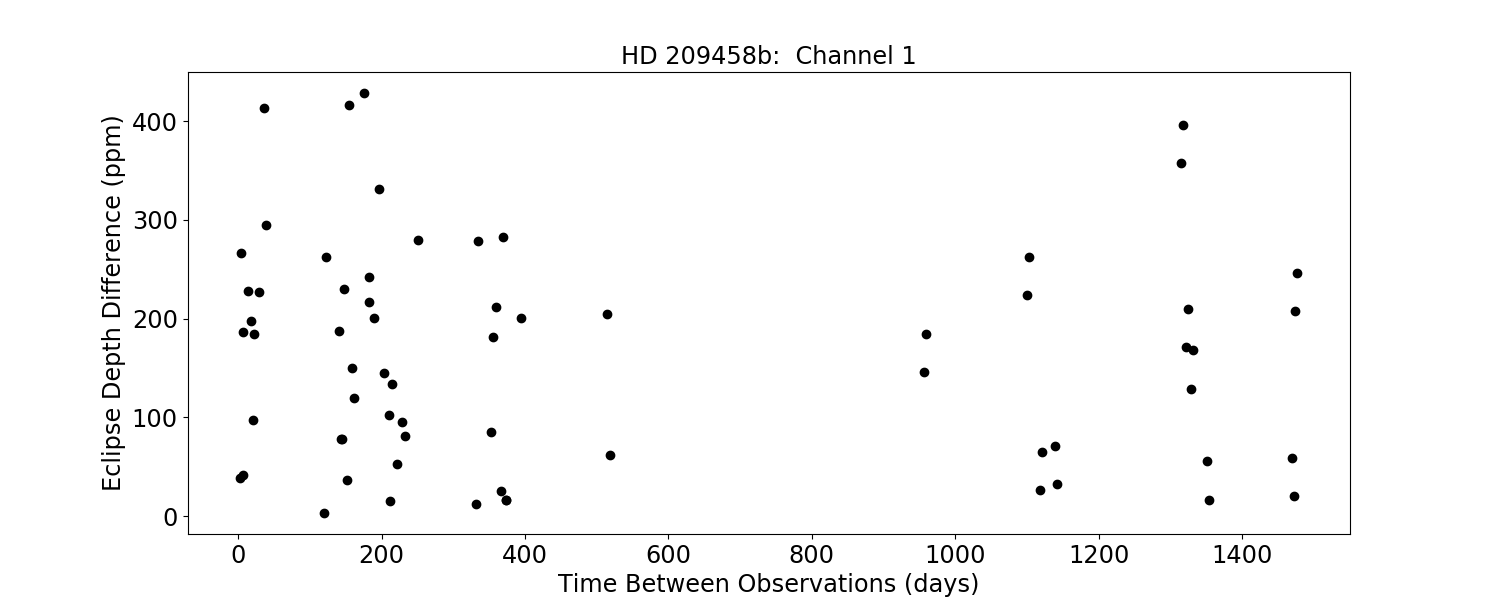}\\
\includegraphics[trim=0.0in 0.0in 0.0in 0.in,clip, width=0.45\textwidth]{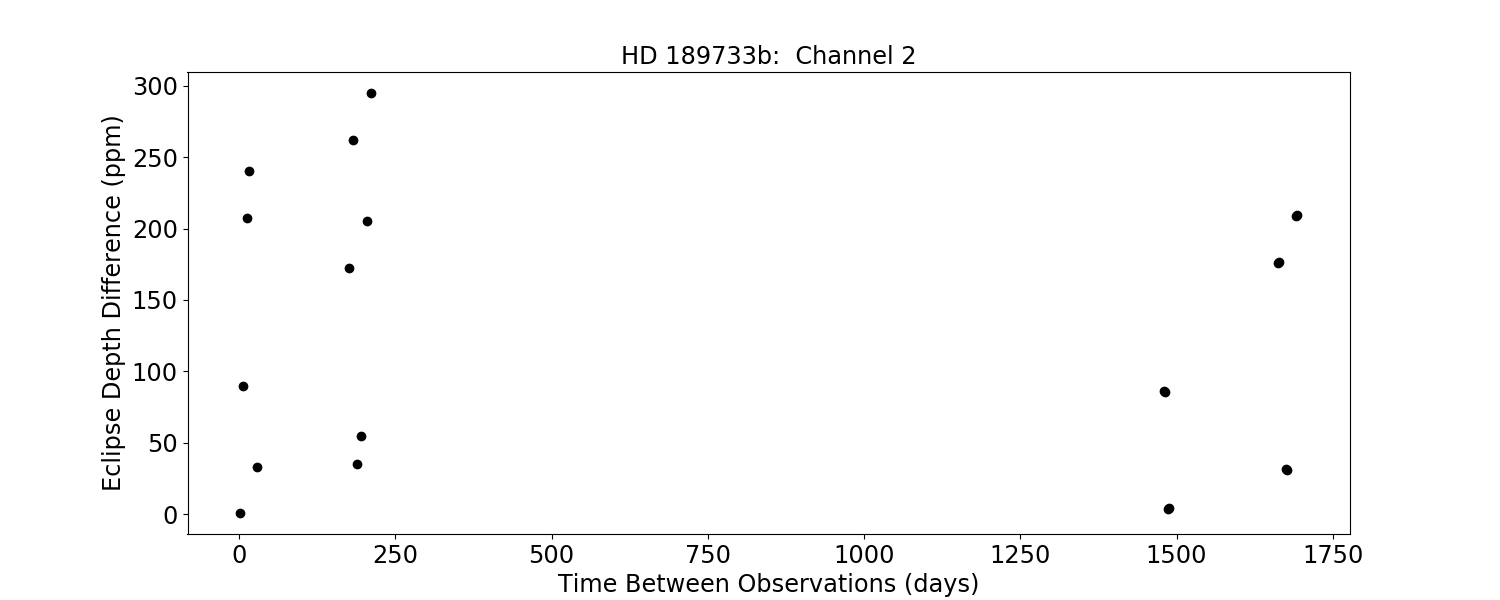}\\
\includegraphics[trim=0.0in 0.0in 0.0in 0.0in,clip, width=0.45\textwidth]{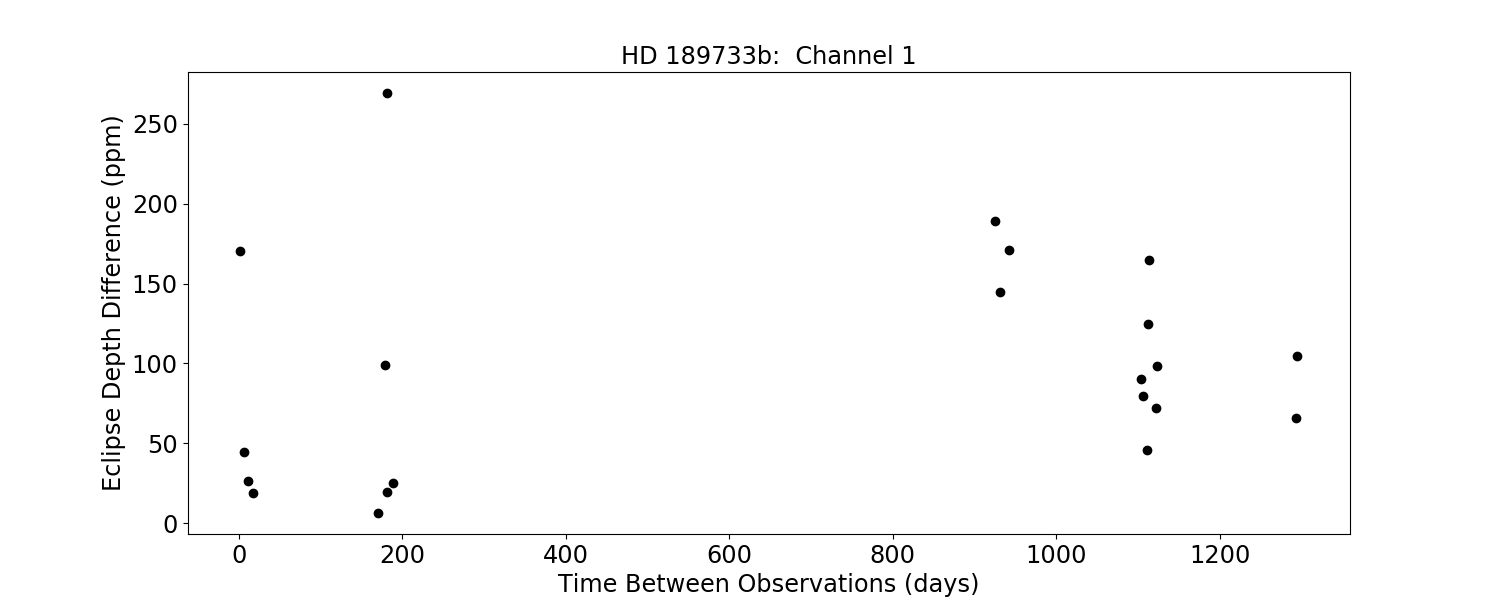}\\

\caption{The absolute difference in eclipse depth for each pair of observations are plotted as a function of the time between observations. Any periodic variability would manifest as peaks in this plot.  The data appears to be randomly scattered suggesting that any variability in the eclipse depths is likely due to precision limitations rather than any periodic variability in the astrophysical signal. }
 \label{time_corr_dep_209}
 \end{figure}

 As a probe of periodic variability in the eclipse depths we calculate the absolute difference in eclipse depths for each pair of observations.  The absolute difference is plotted against the time between observations.  Small numbers of observations do not lend themselves to more thorough methods of detecting power at certain time scales such as Lomb-Scargle periodograms, however, this simple approach would show a spike in absolute difference at any relevant timescale.  The apparent random scatter of the data points in Figure \ref{time_corr_dep_209} does not indicate any periodic structure in any of the observations. Given that we find no evidence for variability in our observations we compare the standard deviation and the magnitude of the  eclipse depths to constrain variability to less than (5.6\% and 6.0\%) and  (12\% and 1.6\%) for channels (1,2) of HD 189733b and HD 209458b respectively. 
 We  synthesized a periodic signal over a grid of varying periods and amplitudes to demonstrate that our observations would be sensitive to variability above the levels of our constraint.  We then simulated measurements using the relevant uncertainty at each of the observation times.  We evaluate the delta BIC for a constant signal versus a periodic signal at each grid point averaged over 100 trials. We find that the detection threshold is not sensitive to period over the timescales probed by these observations; only amplitude.  The delta BIC is noisy and insignificant until amplitudes equivalent to our previously stated constraints after which it increases rapidly with signal amplitude.  This metric provides evidence that if a signal were present our observations would be sensitive to it.

\subsection{Stellar Variability}
Stellar variability in the form of spots or plages can affect the measured planetary transit signal, potentially altering one's interpretation of its atmospheric composition, particularly for bright targets with large transit signals \citep{pont08, silvavalio08, czesla09, wolter09, agol10, berta11, carter11, desert11, sing11, fraine14, mccullough14, oshagh14, damasso15, barstow15, zellem15, zellem17, cauley17, cauley18, rackham17, rackham18, rackham19, morris18}. HD 189733 is an active K0 star which has been observed to
vary by as much as $\pm$ 1.5\% at visible wavelengths with a rotation period of 11.95 days \citep[][and references within]{Knutson2012} whose variability potentially impacts the observed transit depth of the planet \citep{mccullough14,Knutson2012}. Although the
amplitudes of these variations are reduced as our Spitzer/IRAC observations are in the infrared \citep{oshagh14, rackham17, morris18} and of eclipses \citep{zellem17}, we examine the amplitude of these changes in comparison to the scatter and uncertainty in our eclipse depth measurements.

\begin{figure}[!t]
\vspace{0.1in}
 \centering

\includegraphics[trim=0.0in 0.0in 0.0in 0.25in,clip, width=0.45\textwidth]{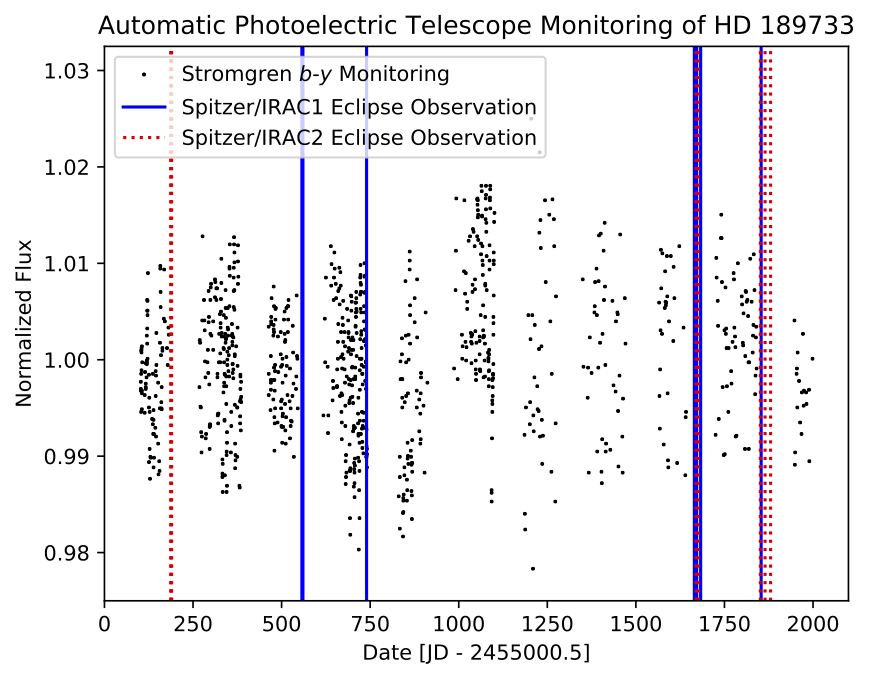}\\
\includegraphics[trim=0.0in 0.0in 0.0in 0.25in,clip, width=0.45\textwidth]{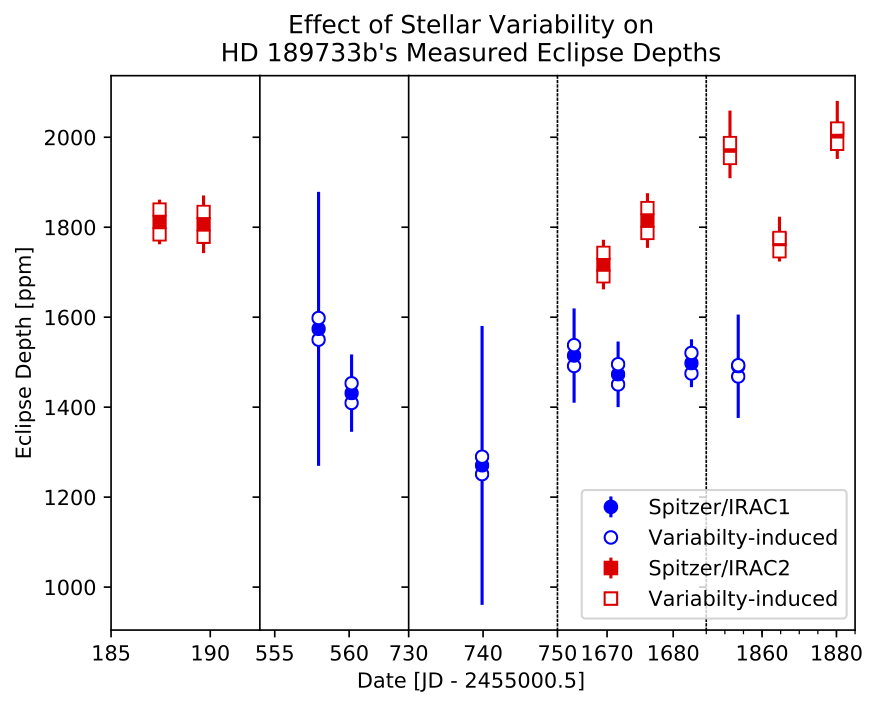}
\caption{$Top:$ HD~189733's stellar variability from the 2009--2014 observing seasons with our \Spitzer/IRAC observations also indicated (IRAC1 in solid blue and IRAC2 in dotted red). $Bottom:$ Conservatively using HD~189733's overall peak-to-trough variability ($Top$), we estimate its impact on the planet's eclipse depth using the prescriptions described in \citet{Knutson2012} and \citet{zellem17}. We find that any variability-induced changes to HD~189733b's eclipse depths (empty blue diamonds for IRAC2 and empty red squares for IRAC2) fall within our measurement uncertainties. Therefore, HD~189733's stellar variability does not statistically impact our measured eclipse depths.}
 \label{fig: stel_var}
 \vspace{0.1in}
 \end{figure}

We employ ground-based monitoring spanning the 2009 to 2014 observing seasons with the Tennessee State University 0.8~m Automated Photoelectric Telescope (APT) at Fairborn Observatory in southern Arizona \citep[e.g.,][Fig.~\ref{fig: stel_var}]{henry99, eaton03}. However, most of our \Spitzer/IRAC observations occur outside the APT observing season (Fig.~\ref{fig: stel_var}). While previous studies have interpolated APT monitoring to their \Spitzer observations \citep{Knutson2012} using an activity model \citep{aigrain12} to fit the APT data and interpolated it to their \Spitzer observations, we conservatively take the peak-to-trough Str{\"o}mgren $b$+$y$ variability observed by APT from 2009 to 2014 (4.7\%) and interpolate it to the infrared using the scaling presented in \citet{Knutson2012} (1.6\%). Using Equation~7 in \citet{zellem17}, we estimate the effect of HD~189733's variability on the observed eclipse depths and find that variability does not statistically impact HD~189733b's eclipse depths as the changes in its measured eclipse depths induced by variability are less than the measurement uncertainties of the eclipses themselves (Fig.~\ref{fig: stel_var}).

\section{Discussion}\label{sec: Dis}

\subsection{Model Predictions}
Given that our assessment of atmospheric variability is in direct relation to the planet's atmospheric dynamics, we compare our {\it Spitzer} variability estimates to predictions from three-dimensional general circulation models (GCMs).  In particular, we use GCM simulations of HD 189733b from \citet{Showman2009} and HD 209458b from \citet{2016Kataria}, respectively, which utilize the Substellar and Planetary Radiation and Circulation (SPARC) model \citep{Showman2009}. The SPARC model couples the general circulation model maintained at the Massachusetts Institute of Technology (the MITgcm, \cite{2004Adcroft}) with a plane-parallel, two-stream version of the multi-stream radiation code developed by \citet{1999MarleyMc}. Further details on these models are provided in \citet{Showman2009}), \citet{2016Kataria} and references therein.
 
\begin{figure}[!ht]
 \centering
 \vspace{0.1in}
\includegraphics[trim=0.0in 0.0in 0.0in 1.0in,clip, width=0.45\textwidth]{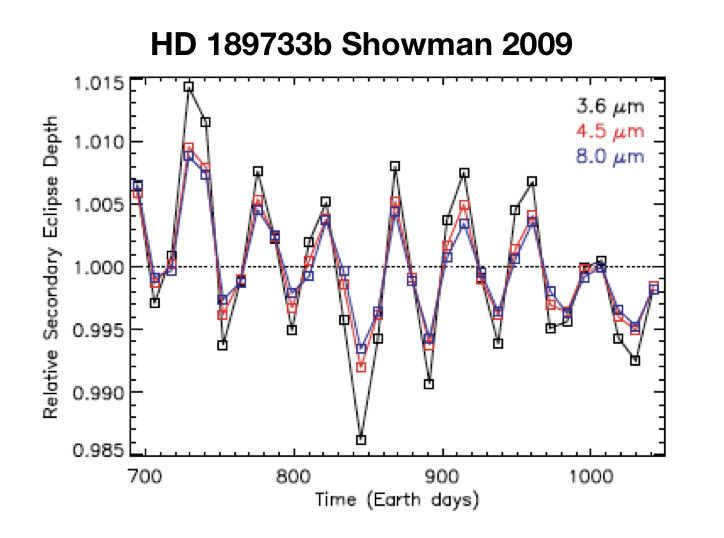}
\includegraphics[trim=0.0in 3.0in 0.0in 1.0in,clip, width=0.45\textwidth]{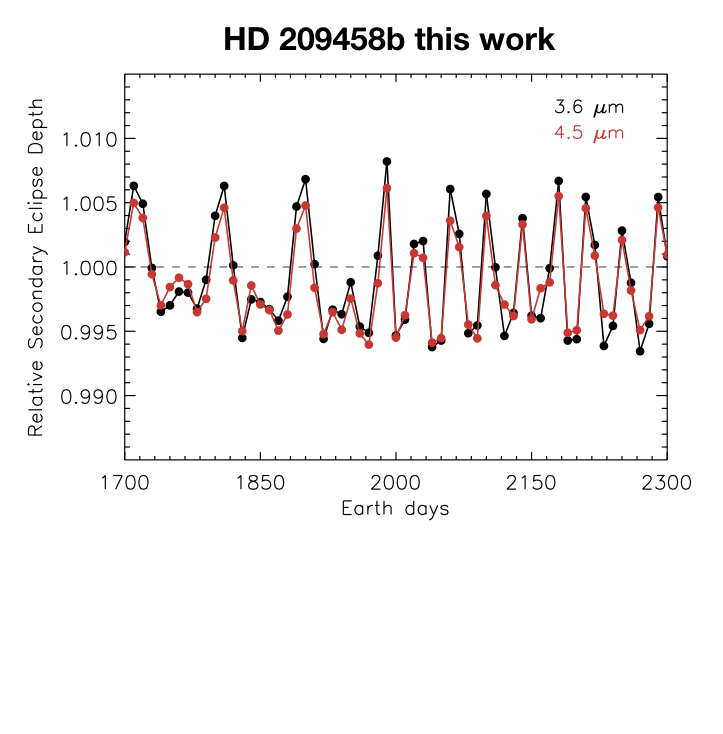}
\caption{Model predictions of variability in eclipse depth orbit to orbit. The top panel is taken from \cite{Showman2009} and shows the predicted change in eclipse depth orbit to orbit for HD 189733b at 3.6, 4.5, and 8.0 \micron.  The bottom panel shows the predicted variation for HD 209458b at 3.6 and 4.5 \micron. In both cases the eclipse depths are scaled by the mean eclipse depth to illustrate only the relative change.}
 \label{fig: depth_predictions}
 \vspace{0.1in}
 \end{figure}

Using the prescription described in \citet{2006Fort}) and \citet{Showman2009}, we compute simulated eclipse depths derived from our GCM results for each planet over a period of 1-2 Earth years (Figure \ref{fig: depth_predictions}).  At each {\Spitzer} bandpass, the simulations, on average, exhibit a periodic variation in eclipse depth of 1-1.5 \% over a period of $\sim$43 days for HD 189733b and a variation of 0.5-1 \% over a period of $\sim$40 days for HD 209458b.  This predicted variability is the result of a global sloshing mode or wave dynamics deep in the planetary atmosphere that leave only a small measurable effect at observable pressures \citet{Showman2009}.  Therefore, hot Jupiters such as these should exhibit low ($\sim$1\%) variability over timescales much longer than the planet's orbital period and smaller than the typical uncertainty in eclipse depth with current instruments.

The low observed atmospheric variability of HD~189733b and HD~209458b with \Spitzer is in contrast to observations of brown dwarfs and solar system gas giants where temporal variability at infrared wavelengths has been observed at the 10-50\% level \citep[see review by][]{artigau18}.  The low level of observable infrared variability for hot Jupiters like HD~189733b and HD~209458b compared with brown dwarfs and solar system gas giants likely arises from differences in the relative strengths in radiative and advective processes taking place in their atmospheres. Because brown dwarfs are self-luminous and comparatively fast rotators ($T_{orb} \sim $hrs), their circulation is dominated by multiple bands of jets and vortices (Showman et al. 2019).  In contrast, the high stellar insolation and synchronous (and hence slower) rotation of hot Jupiters results in the emergence of strong day-night forcing that produce fast ($\sim$1 km/s) planetary-scale jets.  The strong radiative forcing and global-scale circulation patterns in the observable portion of hot Jupiter atmospheres suppress small scale variations in the planet's thermochemical structure that would contribute to large-amplitude variations in the dayside flux from the planet.


 \subsection{Further Observational Probes of Variability}

Our ability to probe variability in exoplanet atmospheres is currently limited by the available targets and observational facilities. It is important to remember that facilities like {\it Spitzer} were not originally designed for high-precision time-series observations. With hot Jupiter targets that orbit bright nearby host stars like HD~209458b and HD~189733b; per eclipse precisions on the order of 100~ppm can be achieved with {\it Spitzer}.  Future observational facilities, such as \JWST, will both improve the achievable precision on eclipse depth measurements and greatly expand the wavelengths over which they can be obtained \citep[][]{beichman14}. Hot Jupiter models predict significant variations in temperature, chemistry, and circulation patterns as a function of pressure level in the atmosphere  \citep[e.g.][]{Showman2009}, which translates into a strong wavelength dependence in the predicted levels of observable variability \citep[e.g.][]{2014Lewis}.  The observations presented here at 3.6 and 4.5 $\mu$m for HD~209458b and HD~189733b are predicted to probe a fairly limited pressure range in the $\sim$1-100~mbar level of these planets atmospheres \citep[e.g.][]{Showman2009}.
Future observations of hot Jupiters at higher precision spanning a larger range of wavelengths as well as observations of cooler exoplanets will be critical for expanding our understanding of the physical processes driving variability, or the lack thereof, in exoplanet atmospheres.

 \begin{figure}[!t]
 \centering
\includegraphics[trim=0.0in 0.0in 0.0in 0.0in,clip, width=0.45\textwidth]{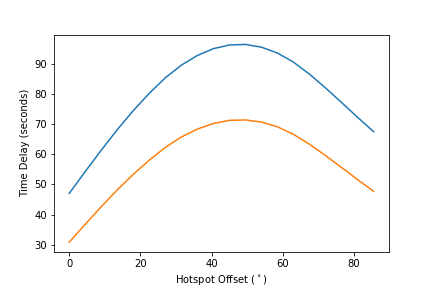}
\caption{Apparent delay in center of eclipse time as a function of the offset of the hot spot for HD 209458b (blue) and HD 189733b (orange) at 4.5 $\mu$m. An offset hot spot causes changes to the shape of ingress/egress that will manifest as a delay in the eclipse timing when being fit with a uniform occultation model.  }
 \label{offset_delay}
 \end{figure}

\subsection{Non-Uniform Disc}

There is prior evidence of non-uniform thermal distributions in both of the targets studied here.  The analysis of the full orbit phase curves of HD 189733b \citep{Knutson2012} provide evidence for an eastward shift of the hottest portion of the planetary atmosphere with respect to the substellar point of 5.29 $\pm$ 0.59 and 2.98 $\pm$ 0.82 hours in channels 1 and 2 respectively.  These offsets correspond to longitudinal offsets of 35$^\circ$ $\pm$ 4$^\circ$ and 20$^\circ$ $\pm$ 6$^\circ$.  Similarly, phase curve analysis of HD 209458b at 4.5 $\mu$m by \citet{2014Zellem} indicates an offset of 41$^\circ$ $\pm$ 6$^\circ$.

\citet{2006Williams} and \citet{2007Rauscher} predicted that a non-uniform dayside temperature distribution will change the shape of ingress/egress in comparison to a uniformly bright model.   
Also, as \citet{2006Williams} predicted and \citet{Agol2010} observed, fitting the eclipse of a planet with a non-uniform dayside temperature distribution with a uniform occultation model will result in an apparent delay in the eclipse time as a result of these changes to the shape of ingress/egress when the timing of the eclipse is a free parameter.  
In Figures \ref{figure:alldepths1} and \ref{figure:alldepths2} we plot the difference between the center of eclipse time and what would be expected by assuming that the eclipse would occur at exactly one half of a period from the transit.  HD 209458b consistently shows an offset in the eclipse time of 155 $\pm$ 25 seconds in Channel 1 and 129 $\pm$ 16 seconds in Channel 2.  HD 189733b shows an offset of 28 $\pm$ 27 seconds in Channel 1 and 35 $\pm$ 13 seconds in Channel 2.

Figure \ref{offset_delay} shows the magnitude of the effect of longitudinal offset on the observed center of eclipse time.  We create a model planet with a hotspot of radius (0.5 R$_{\rm planet}$) utilizing the python package SPIDERMAN \citep{2017ascl.soft11019L}.  The global (out of hot spot) temperature is set to 1200 K and the hot spot temperature to 1700 K. Starting with the hot spot centered at the substellar point, we move the hotspot longitudinally in 1$^\circ$ increments and fit at each iteration with a uniform brightness model with center of eclipse time as a free parameter.  We plot in Figure \ref{offset_delay} the delay in eclipse time relative to the predicted time exactly one half period from transit.  Note that some portion of apparent delay is present even in the 0$^\circ$ offset case.  This is due to light travel time across the solar system ($\sim$ 45 sec. for HD 209).  The offset of the maximum brightness observed in the phase curves would suggest that eclipse observations fit with a uniform model should exhibit an offset in eclipse timing of 40 and 60 seconds (Ch. 1 and 2)  for HD 189733b and 90 seconds for HD 209458b channel 2 observations.  The secondary eclipse fits presented here show strong agreement with the phase curve observations of HD 209458b. We are unable to confirm the phase curve offsets for HD 189733b with the precision attained with this set of observations.  

\begin{figure}[!t]
 \centering
 \vspace{0.2in}
\includegraphics[trim=1.0in 2.0in 1.0in 2.25in,clip, width=0.45\textwidth]{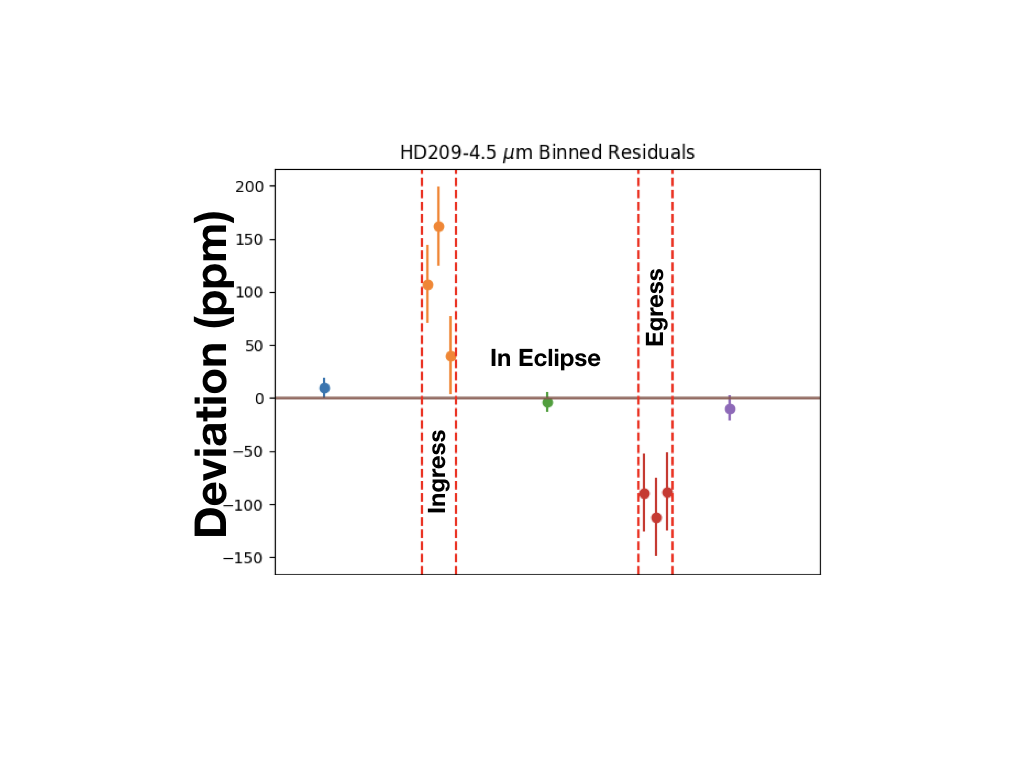}
\caption{Here we stack the residuals from the best fit of each  HD 209458b channel 2 observation with the time of mid eclipse fixed to coincide with phase 0.5 plus the transit time of light across the system.  We bin them into 9 separate bins: one each for pre-eclipse, during eclipse, and post-eclipse and three each for ingress and egress.  The structure shown in ingress/egress is similar to model predictions illustrated in \citet{2006Williams, 2007Rauscher} resulting from fitting a non-uniform dayside temperature distribution with a uniform occultation model.} 
 \label{residuals}
 \end{figure}

As a further probe of non-uniformity, we refit the data while fixing the eclipse time to the expected value based on previous observations of epoch of transit and assuming a circular orbit.  We combine the residuals of all the fits of each target in each channel to probe for any coherent deviations from the uniform model.  
Fig. \ref{residuals} shows the stacked residuals from all of the channel 2 fits of HD 209.  Here we see strong evidence for deviations in the shape of ingress/egress that could be caused by an eastward shifted hotspot.  There are other factors that could also mimic this effect.  The effect of eccentricity and shape of planet are explored in detail by \citet{2012Dewit}.  However, the precision with which the orbital properties of both of these canonical hot-Jupiters have been characterized allow many of factors to be ruled out or tight enough constraints to be placed on them to make a significant detection of non-uniformity likely in future eclipse mapping studies \citep{2018Rauscher}.   

 \section{Conclusion}

This analysis of multi - epoch secondary eclipse observations of the canonical hot Jupiters, HD 189733b and HD 209458b,  finds no evidence for temporal variability greater than the precision of our observations.   We expect variability in secondary eclipse depth to be driven by the dynamics of the atmosphere thus motivating comparison of observations to GCM predictions. The simulations, on average, exhibit a periodic variation in eclipse depth of 1-1.5 \% for both planets at both 3.6 and 4.5 $\mu$m.  Based on the uncertainty in each measurement of secondary eclipse depth, we are able to constrain variability to less than (5.6\% and 6.0\%) and  (12\% and 1.6\%) for channels (1,2) of HD 189733b and HD 209458b respectively.  The lack of evidence of variability provides motivation and justification to combine multi - epoch observations to achieve the precision necessary to perform high spatial resolution techniques such as eclipse mapping.  The apparent offset of the center of eclipse time evident in these observations provides further evidence for a non-uniform dayside temperature distribution for the planet HD 209458b.  We compare the evidence for non - uniformity to previous analyses of full phase curve observations and find agreement with HD 209458b observations at 4.5 \micron. The precision achieved by these observations when phase folding and stacking all like observations is enough to resolve deviations to ingress/egress caused by a non - uniform temperature distribution in at least the case of HD 209458b at 4.5 $\mu$m.  This level of precision meets the requirements to utilize the eclipse mapping technique to produce spatially resolved, two dimensional maps of the planetary dayside thermal distribution.  
 
 \section{Acknowledgements}
 
 This work was supported by NASA Headquarters under the NASA Earth and Space Science Fellowship Program under Grant Number 80NSSC17K0484.  This work is based on observations made with the {\sl Spitzer Space Telescope}, which is operated by the Jet Propulsion Laboratory, California Institute of Technology under a contract with NASA.  We also acknowledge that part of this work was completed at the Space Telescope Science Institute (STScI) operated by AURA, Inc.  This research has made use of the Exoplanet Orbit Database
and the Exoplanet Data Explorer at exoplanets.org.  
 
\bibliography{variabilitybib}

\end{document}